\documentclass[aps,prl,reprint,groupedaddress]{revtex4-1}
\usepackage{graphicx} % Include figure files
\usepackage{dcolumn} % Align table columns on decimal point
\usepackage{bm}% bold math
\usepackage{hyperref}% add hypertext capabilities

\begin{document}

\title{ Potential energy surface and formation of superheavy nuclei with the Skyrme energy-density functional }
\author{Cheng Peng }
\author{Zhao-Qing Feng }
\email{Corresponding author: fengzhq@scut.edu.cn }

\affiliation{School of Physics and Optoelectronics, South China University of Technology, Guangzhou 510640, China}

\date{\today}

\begin{abstract}

With the Skyrme energy-density functional theory, the nucleus-nucleus potential is calculated and potential energy surface is obtained with different effective forces for accurately estimating the formation cross sections of superheavy nuclei in massive fusion reactions. The width and height of the potential pocket are influenced by the Skyrme effective forces SkM, SkM$^{\ast}$, SkP, SIII, Ska and SLy4, which correspond to the different equation of state for the isospin symmetry nuclear matter. It is found that the nucleus-nucleus potential is associated with the collision orientation and Skyrme forces. More repulsive nuclear potential is pronounced with increasing the incompressible modulus of nuclear matter, which hinders the formation of superheavy nuclei. The available data in the fusion-evaporation reaction of $^{48}$Ca+$^{238}$U are nicely reproduced with the SkM$^{\ast}$ parameter by implementing the potential into the dinuclear system model.

\begin{description}
\item[PACS number(s)]
21.30.Fe, 24.60.-k, 25.70.Jj        
\end{description}
\end{abstract}

\maketitle

\section{I. Introduction}

The heavy-ion fusion reactions attracted much attention since the first experiments at Crocker Laboratory with cyclotron accelerator in 1950s \cite{Mi50}. The multidimensional quantum tunneling, collective excitation, nucleon or cluster transfer etc, influence the fusion dynamics and fusion probability \cite{Ba98}. Up to now, the nuclear fusion reactions have been extensively investigated, in particular on the topics of weakly bound nuclei induced reactions \cite{Ca06}, nuclear fusion at deep sub-barrier energies for astrophysical interests \cite{Ad11} and synthesis of superheavy nuclei (SHN) \cite{1Sh,Og15}. Roughly, one third nuclides on the nuclear chart were synthesized in laboratories via the fusion reactions \cite{Th13}. The hunting of superheavy nuclei (SHN) in the nature or synthesizing SHN in laboratories, in particular around the 'island of stability' predicted theoretically, is the topical issue in past and nowadays. The cold-fusion reactions with $^{208}$Pb or $^{209}$Bi based targets were firstly proposed by Oganessian et al. \cite{Og75}. The superheavy elements (SHEs) from Bh to Cn were successfully synthesized in the cold-fusion reactions at GSI (Darmstadt, Germany) with the heavy-ion accelerator UNILAC and the SHIP separator \cite{Ho00,Mu15}. Experiments on the synthesis of element Nh (Z=113) in the $^{70}$Zn+$^{209}$Bi reaction have been performed successfully at RIKEN (Tokyo, Japan) \cite{Mo04}. The SHEs from Fl (Z=114) to Og (Z=118) have been synthesized at the Flerov Laboratory of Nuclear Reactions (FLNR) at Dubna (Russia) with the double- magic nuclide $^{48}$Ca bombarding actinide nuclei \cite{Og99,Og06}. With constructing the new facilities in the world such as RIBF (RIKEN, Japan) \cite{Sa18}, SPIRAL2 (GANIL in Caen, France) \cite{Ga10}, FRIB (MSU, USA) \cite{msu}, HIAF (IMP in Huizhou, China) \cite{Ya13}, the SHNs on the 'island of stability' by using the neutron-rich radioactive beams induced fusion reactions or via the multinucleon transfer (MNT) reactions might be possible in experiments. Sophisticated models are expected for understanding the nuclear dynamics of SHN formation close to the 'island of stability' via the massive fusion reactions or MNT mechanism, i.e., the quasi-fission dynamics, fusion-fission, incomplete and complete fusion reactions, preequilibrium cluster emission, deep-inelastic collisions etc.

The fusion dynamics is governed by the nucleus-nucleus (NN) interaction potential, which is estimated with the frozen density or time dependent density profile of colliding system. Recently, it has been found that the Pauli exclusion principle is of significance in the nuclear potential and influences the width of the potential pocket \cite{Sc17,Vo18,Ua21}. An empirical formula was proposed by Bass for estimating the Coulomb barrier and fusion cross section \cite{Ba74}. In the light and medium reaction systems, the compound nucleus is formed after overcoming the Coulomb barrier. However, the quasifission mechanism appears in the heavy colliding systems, in which the disintegration of colliding system after a few of nucleon transfer hinders the compound nucleus formation. The neck dynamics, shape evolution, collective excitation and nucleon transfer influences the NN potential. It has been known that the NN potential in the fusion reactions is associated with the shape evolution and beam energy by the dynamical models, e.g., time-dependent Hartree-Fock (TDHF) approach \cite{Si18,AS10}, quantum molecular dynamics (QMD) model \cite{Fe05}. There are mainly two sorts of NN potential, namely, the phenomenological potentials such as the Woods-Saxon potential \cite{Wo73, Ch76}, proximity potential \cite{My00}, potentials (Yukawa-plus-exponential, DDM3Y, Migdal etc) via the double-folding method \cite{Kr79,Kh93,Sp77,Ad96}, and the adiabatic potential \cite{Za07}. It is also possible to construct the NN potential within the energy-density functional approach based on the effective nucleon-nucleon interaction, i.e., the Skyrme force \cite{Sk56,Br85,BR75,BR78}, the finite-range Gogny interaction \cite{Go80} etc. The advantage of the energy-density functional approach establishes a unified description of nuclear structure, nuclear dynamics and nuclear matter based on the effective nucleon-nucleon interaction.

In this work, the NN potential is calculated within the Skyrme energy-density functional. The potential energy surface is obtained with the approach and the production of SHN is discussed by implementing into the dinuclear system model. The article is organized as follows. In  Section 2 we give a brief description of Skyrme energy-density functional theory and the NN potential. The driving potential and SHN production in the reaction of $^{48}$Ca+$^{238}$U are shown In Section 3. A summary and perspective on the NN potential from the microscopical method are presented in Section 4.

\section{II. Brief description of the model}
\subsection{2.1  Skyrme energy density functional and nucleus-nucleus potential }

The nucleus-nucleus potential is a basic quantity for describing the nuclear dynamics in the low-energy heavy-ion collisions. The interaction potential in binary collisions depend on the collision orientation and is composed of nuclear and Coulomb contribution as follows \cite{14zq,15zq}
\begin{eqnarray}
V(R, {\alpha}_{P}, {\alpha}_{T}, J) && =V_{nucl}(R, {\alpha}_{P}, {\alpha}_{T}) + V_{Coul}(R, {\alpha}_{P}, {\alpha}_{T})              \nonumber \\
&& + \frac{\hbar^{2}}{2\xi_{rel}}J(J+1).
\end{eqnarray}
Here the Coulomb potential is calculated by the Wong's formula \cite{Wo73}. The ${\alpha}_{i}$ denotes the symbols $R_{i}, \theta_{i}, \beta_{i}$ with $i=P, T$ being the projectile or target nucleus and the relative momentum of inertia. The $R_{i}, \theta_{i}, \beta_{i}$ represent the nuclear radii, quadrupole deformations and polar angles between the beam direction and the symmetry axes of deformed nuclei, respectively. The $R$ is the center-of-mass distance of projectile and target nuclides. Shown in Fig. 1 is the definition of the quantities $R_{1}, R_{2}, \theta_{1}, \theta_{2}$ and the integration variables $r, \theta$. The deformation effect is included in the nuclear and Coulomb potentials, which results in the orientation dependence of the Coulomb barrier and influences the quasifission dynamics in the massive fusion reactions. The multiple integral with the energy-density functional by the Skyrme force is performed in the spherical coordinate system ($r$, $\theta$ and $\phi$).

%%%%%%%%%%%%%%%%%%%%%%%%%%%%%%%%%%%% figure 1 %%%%%%%%%%%%%%%%%%%%%%
\begin{figure*}
	\includegraphics[width=16 cm]{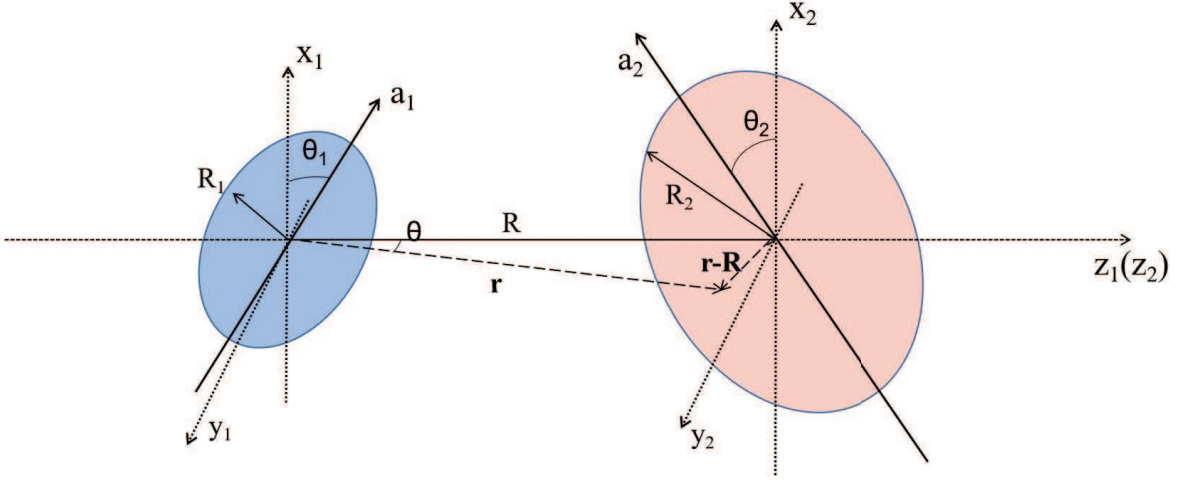}
	\caption{Schematic picture of relative motion in two colliding nuclei. }
\end{figure*}
%%%%%%%%%%%%%%%%%%%%%%%%%%%%%%%%%%%%%%%%%%%%%%%%%%%%%%%%%%%%%%%

The nuclear potential is calculated by the Skyrme energy-density functional as \cite{Ba02,De02}
\begin{eqnarray}
V_{nucl}(R, \{\alpha\}_{P}, \{\alpha\}_{T})  && = E_{sys}(R, \{\alpha\}_{P}, \{\alpha\}_{T})            \nonumber \\
&&  - E_{P}(\{\alpha\}_{P}) - E_{T}(\{\alpha\}_{T}).
\end{eqnarray}
The $E_{sys}$, $E_{P}$ and $E_{T}$ are the binding energies contributed from the nucleon-nucleon force of colliding system, projectile and target nuclei by the relation of $E=\int \int\int \varepsilon[\rho_{p}(\bm{r}), \rho_{n}(\bm{r})]r^{2}\sin(\theta)drd\theta d\phi$, respectively. The energy-density functional $\varepsilon[\rho_{p}(\bm{r}),\rho_{n}(\bm{r})]$ is derived from the Skyrme force as (see in detail in Appendix)
\begin{eqnarray}
 && \hat{V}_{eff}(\bm{r}_{1},\bm{r}_{2})  = t_{0}(1+x_{0}\hat{P}_{\sigma})\delta(\bm{r}_{1}-\bm{r}_{2})                \nonumber \\
 && + \frac{t_{3}}{6}(1+x_{3}\hat{P}_{\sigma})\delta\left( \frac{1}{2}(\bm{r}_{1}+\bm{r}_{2})\right) \delta(\bm{r}_{1}-\bm{r}_{2})         \nonumber \\
 && - \frac{t_{1}}{2}(1+x_{1}\hat{P}_{\sigma})\left( (\bigtriangledown_{1}-\bigtriangledown_{2})^{2}\delta(\bm{r}_{1}-\bm{r}_{2}) + h.c.   \right)     \nonumber \\
 && -t_{2}(1+x_{2}\hat{P}_{\sigma})  \left((\bigtriangledown_{1}-\bigtriangledown_{2})\delta(\bm{r}_{1}-\bm{r}_{2}) (\bigtriangledown_{1}-\bigtriangledown_{2}) \right)          \nonumber \\
 && +t_{2}\hat{\bm{k}}^{2}\cdot \delta(\bm{r}_{1}-\bm{r}_{2}) \hat{\bm{k}}            \nonumber \\
 && +iW_{0}(\hat{\sigma}_{1}+\hat{\sigma}_{2})\cdot  \hat{\bm{k}}^{'}\times \delta(\bm{r}_{1}-\bm{r}_{2})\hat{\bm{k}}
\end{eqnarray}
with the spin-exchange operator
\begin{eqnarray}
 \hat{P}_{\sigma}=\frac{1}{2}(1+\hat{\sigma}_{1}\cdot\hat{\sigma}_{2}),
\end{eqnarray}
\begin{eqnarray}
\hat{\bm{k}}=\frac{1}{2i}(\bigtriangledown_{1}-\bigtriangledown_{2}),\quad \hat{\bm{k}}^{'}=-\frac{1}{2i}(\bigtriangledown_{1}-\bigtriangledown_{2}).
\end{eqnarray}
The zero-range effective forces between nucleons in the nuclear environment provide the energy-density functional and are available for the ground-state properties of finite nuclei and nuclear matter at saturation density \cite{St07}.

The energies of colliding system, projectile and target nuclei are calculated by
\begin{eqnarray}
E_{sys}(R, \{\alpha\}_{P}, \{\alpha\}_{T}) && =  \int\varepsilon[\rho_{1p}(\bm{r})+\rho_{2p}(\bm{R-r}),         \nonumber \\
	&&\rho_{1n}(\bm{r})+\rho_{2n}(\bm{R-r}) ] d\bm{r} ,
\end{eqnarray}

\begin{equation}
	 E_{P}(\{\alpha\}_{P})=\int\varepsilon[\rho_{1p}(\bm{r}), \rho_{1n}(\bm{r}) ] d\bm{r}
\end{equation}
and
\begin{equation}
	 E_{T}(\{\alpha\}_{T})=\int\varepsilon[\rho_{2p}(\bm{R-r}), \rho_{2n}(\bm{R-r}) ] d\bm{r},
\end{equation}
respectively. The density profiles of proton and neutron distributions for projectile and target nuclides are taken to be frozen of the Woods-Saxon form as
\begin{equation}
 \rho_{1i}(\bm{r})=\frac{\rho_{01i}}{1+\exp[r-R_{P}/a_{i}]},     i=\{n,p\}
\end{equation}
with the diffuseness coefficients $a_{i}$ being the values of 0.55-0.65 fm and the saturation density $\rho_{01i}$=0.06-0.09 fm$^{-3}$ calculated by the Skyrme-Hartree-Fock method.
The projectile radii with the quadrupole deformation is given by
\begin{eqnarray}
R_{P}=R_{0P}\left( 1+\sqrt{\frac{5}{4\pi}}\beta_{2}\frac{3cos^{2}\theta-1}{2}\right)
\end{eqnarray}
with
$R_{0P}=1.28A_{P}^{1/3}-0.76+0.8A_{P}^{-1/3}$.
Similarly, the neutron and proton density distributions of target nucleus are obtained.

Within the help of the well-known extended Thomas-Fermi (ETF) approximation, the kinetic energy term is obtained up to the second order extension. The energy-density is expressed as \cite{Br85}
\begin{equation}
\varepsilon [\rho_{p}(\bm{r}),\rho_{n}(\bm{r})] = \frac{\hbar^{2}}{2m}[\tau_{p}(\bm{r})+\tau_{n}(\bm{r})] + \nu_{sk}(\bm{r})
\end{equation}
with the kinetic energy term
\begin{eqnarray}
\tau_{i}(\bm{r})=&&\frac{3}{5}(3\pi^{2})^{2/3}\rho_{i}^{5/3}+\frac{1}{36}\frac{(\bigtriangledown\rho_{i})^{2}}{\rho_{i}}+\frac{1}{3}\bigtriangleup\rho_{i} \nonumber \\
	&&+\frac{1}{6}\frac{\bigtriangledown\rho_{i}\bigtriangledown f_{i}+\rho_{i}\bigtriangleup f_{i}}{f_{i}}-\frac{1}{12}\rho_{i} \left(\frac{\bigtriangledown f_{i}}{f_{i}} \right)^{2}              \nonumber \\
	&&+\frac{1}{2}\rho_{i} \left( \frac{2m}{\hbar^{2}}\frac{W_{0}}{2}\frac{\bigtriangledown(\rho+\rho_{i})}{f_{i}} \right)^{2}
\end{eqnarray}
with
\begin{eqnarray}
	f_{i}=1+\frac{2m}{\hbar^{2}}(\frac{3t_{1}+5t_{2}}{16}+\frac{t_{2}x_{2}}{4})\rho_{i}(\bm{r}).
\end{eqnarray}
Here the local density $\rho_{i}$ with $i=n, p$, $\rho = \rho_{n}+\rho_{p}$ and the kinetic energy density $\tau=\tau_{n}+\tau_{p}$ are satisfied in the calculation. The potential part in the energy-density functional is given by
\begin{eqnarray}
	&&\nu_{sk}(\bm{r})=\frac{t_{0}}{2}[(1+\frac{1}{2}x_{0})\rho^{2}-(x_{0}+\frac{1}{2})(\rho_{p}^{2}+\rho_{n}^{2})]    \nonumber \\
	&&+\frac{1}{12}t_{3}\rho^{\alpha}[(1+\frac{1}{2}x_{3})\rho^{2}-(x_{3}+\frac{1}{2})(\rho_{p}^{2}+\rho_{n}^{2})]      \nonumber \\
	&&+\frac{1}{4}[t_{1}(1+\frac{1}{2}x_{1})+t_{2}(1+\frac{1}{2}x_{2})]\tau\rho          \nonumber \\
	&&+\frac{1}{4}[t_{2}(x_{2}+\frac{1}{2})-t_{1}(x_{1}+\frac{1}{2})](\tau_{p}\rho_{p}+\tau_{n}\rho_{n})    \nonumber \\
	&&+\frac{1}{16}[3t_{1}(1+\frac{1}{2}x_{1})-t_{2}(1+\frac{1}{2}x_{2})](\bigtriangledown\rho)^{2}     \nonumber \\
	&&-\frac{1}{16}[3t_{1}(x_{1}+\frac{1}{2})+t_{2}(x_{2}+\frac{1}{2})]((\bigtriangledown\rho_{n})^{2}+(\bigtriangledown\rho_{p})^{2})    \nonumber \\
	&& - \frac{mW_{0}^{2}}{2\hbar^{2}} \left[\frac{\rho_{p}}{f_{p}}(2\bigtriangledown\rho_{p}+\bigtriangledown\rho_{n})^{2}+
	\frac{\rho_{n}}{f_{n}} (2\bigtriangledown\rho_{n}+\bigtriangledown\rho_{p})^{2} \right].    \nonumber \\
\end{eqnarray}
The parameters $t_{0}$, $t_{1}$, $t_{2}$, $t_{3}$, $x_{0}$, $x_{1}$, $x_{2}$, $x_{3}$, the density-dependent stiffness $\alpha$ and the spin-orbit strength $W_{0}$ are listed in Table 1. The six sets of Skyrme parameters SkP \cite{Do84}, SkM, SkM$^{\ast}$ \cite{Kr80}, SLy4 \cite{Ch98}, Ska \cite{Ko76}, SIII \cite{Be75} are taken in the calculation. The binding energy, root-mean-square radii of finite nuclei around the magic numbers and nuclear matter properties at saturation density are self-consistently described with the forces.

%%%%%%%%%%%%%%%%%%%%%%%%%%%%%%%%%%%%%%% table 1 %%%%%%%%%%%%%%%%%%%%%%%%
\begin{table*} [!htb]
	\centering
	\caption{Parameters of the Skyrme forces used in the calculation. }
	\label{tab1}
	\begin{ruledtabular}

           \begin{tabular}{cccccccc}
 & Parameter   & SkP    & SkM    & SkM$^{\ast}$    & SLy4     & Ska       & SIII        \\
 \hline
 & $t_{0}$ (MeV fm$^{3}$) & -2931.7 & -2645 & -2645 & -2488.91 & -1602.78 &-1128.75   \\
 & $t_{1}$ (MeV fm$^{5}$) & 320.6 & 385 & 410  & 486.82 & 570.88 & 395.0    \\
 & $t_{2}$ (MeV fm$^{5}$) & -337.4 & -120 & -135  & -546.39 & -67.70 & -95       \\
 & $t_{3}$ (MeV fm$^{3+3\alpha}$) & 18709 & 15595 & 15595  & 13777 & 8000 & 14000  \\
 & $x_{0}$ &0.292 & 0.09 & 0.09  & 0.834 & -0.02 & 0.45  \\
 & $x_{1}$ & 0.653 & 0 & 0 & -0.344 & 0 & 0   \\
 & $x_{2}$ &-0.537 & 0 & 0  & -1.0 & 0 & 0   \\
 & $x_{3}$ & 0.181& 0 & 0 & 1.354 & -0.286 & 1   \\
 & W$_{0}$ (MeV fm$^{5}$) & 100 & 130 & 130 &  123 & 125 & 120   \\
 & $\alpha$    & 1/6 & 1/6 & 1/6 & 1/6 & 1/3 & 1   \\
 & $K_{\infty}$ (MeV) & 199 & 217  & 217   & 230 & 261 & 352

            \end{tabular}
	\end{ruledtabular}
\end{table*}
%%%%%%%%%%%%%%%%%%%%%%%%%%%%%%%%%%%%%%%%%%%%%%%%%%%%%%%%%%%%%%%%

The nucleus-nucleus potential is of importance in the heavy-ion fusion reactions and determines the height of Coulomb barrier, the width and shape of potential pocket, quasi-fission barrier etc \cite{24ml,25wn}. Consequently, the quasifission yields, fusion-fission products, fusion cross section, isotopic distribution, angular and kinetic energy spectra in the deep inelastic collisions or multinucleon transfer reactions, are influenced by the potential. As a test, the potentials calculated by the energy-density functional and double-folding method with the Migdal force \cite{Ad96} are compared in Fig. 2 in collisions of $^{70}$Zn+$^{208}$Pb (left panel) and $^{138}$Ba + $^{138}$Ba (right panel), respectively. We select the Skyrme forces SkP, SkM, SLy4, Ska and SIII corresponding to the different incompressible modulus of nuclear matter at the normal density. The hard EOS with SIII leads to rapidly increase of the potential by approaching the projectile-target distance because of the repulsive nuclear potential. The Coulomb potential exhibits the repulsive interaction and is obvious in the symmetric system. The nuclear potential (difference of nucleus-nucleus potential and Coulomb potential) with the Migdal force rapidly varies from the attractive to repulsive interaction with decreasing the distance. A wider and deeper pocket is formed in the reaction of $^{70}$Zn+$^{208}$Pb, which is favorable for the compound nucleus formation. It should be mentioned that the self-consistent description of nuclear structure, reaction and matter is established with the Skyrme energy-density functional in comparison with the double-folding method.

The deformation, collective excitation, shape evolution, initial orientation etc influence the nucleus-nucleus potential \cite{Kn4, Nj5, Ai6}. In the realistic nuclear reaction, the density profile varies with the evolution time of colliding system and leads to the complicated NN potential. Two typical approximations are usually used in the reaction models, i.e., sudden approximation and adiabatic approach. In the calculation, the sudden approximation with the frozen nuclear density is used in the potential energy surface and the estimation of SHN production. As a typical reaction system, the $^{48}$Ca induced fusion reactions on the actinide nuclides were chosen for succussfully synthesizing the SHN with Z=112-118 at Dubna. Shown in Fig. 3 is a comparison of the NN potentials calculated with different Skyrme forces and the double-folding method and the initial angle of symmetry axis and collision direction. The Coulomb barriers with the range of 185-205 MeV are obtained with the Skyrme forces and close to the values of proximity potential ($V_{b}$=198.5 MeV) and static barrier ($V_{b}$=187 MeV) by the quantum molecular dynamics model \cite{Fe05}.

%%%%%%%%%%%%%%%%%%%%%%%%%%%%%%%%%%%% figure 2 %%%%%%%%%%%%%%%%%%%%%%
\begin{figure*}
\includegraphics[width=8 cm]{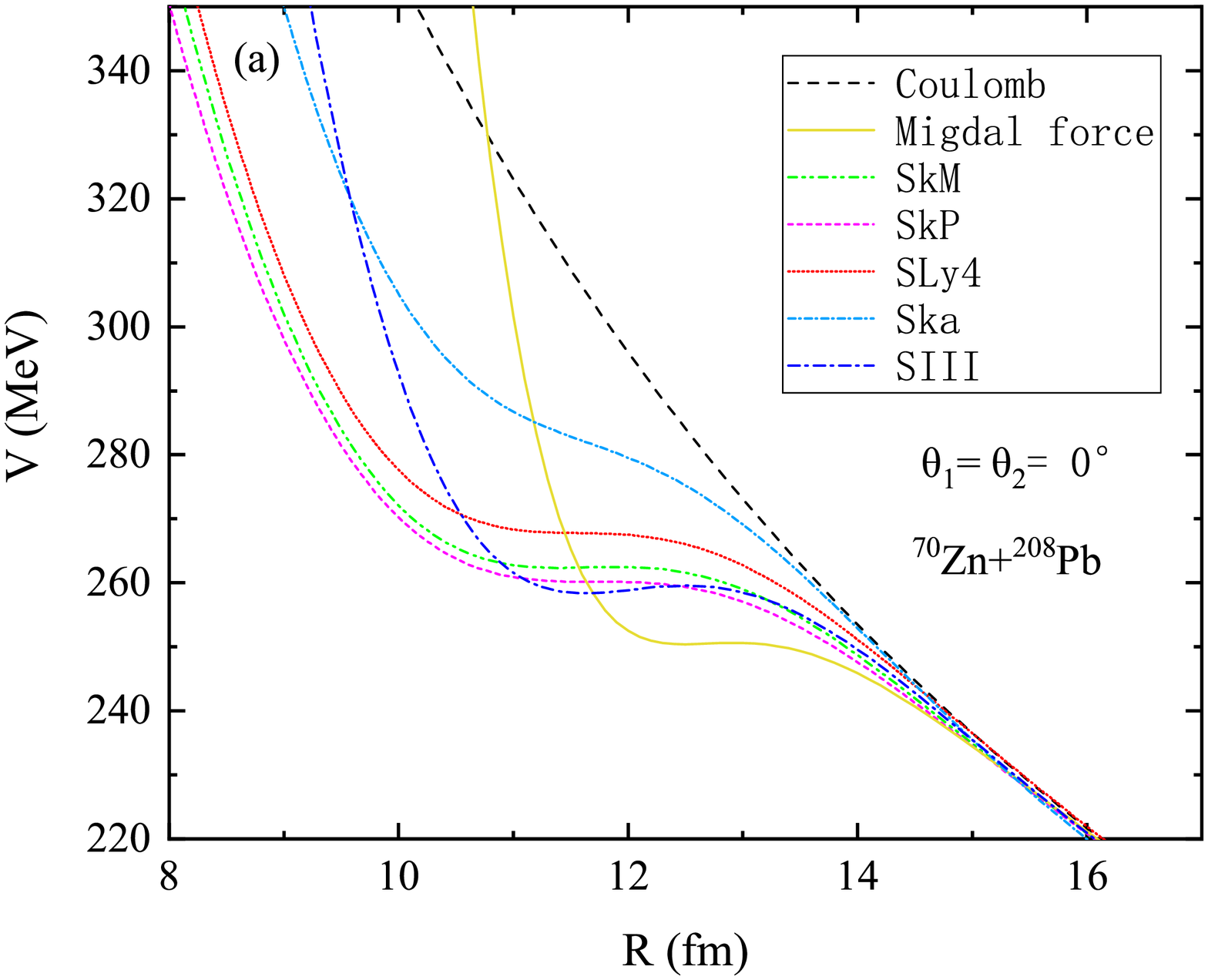} \includegraphics[width=8 cm]{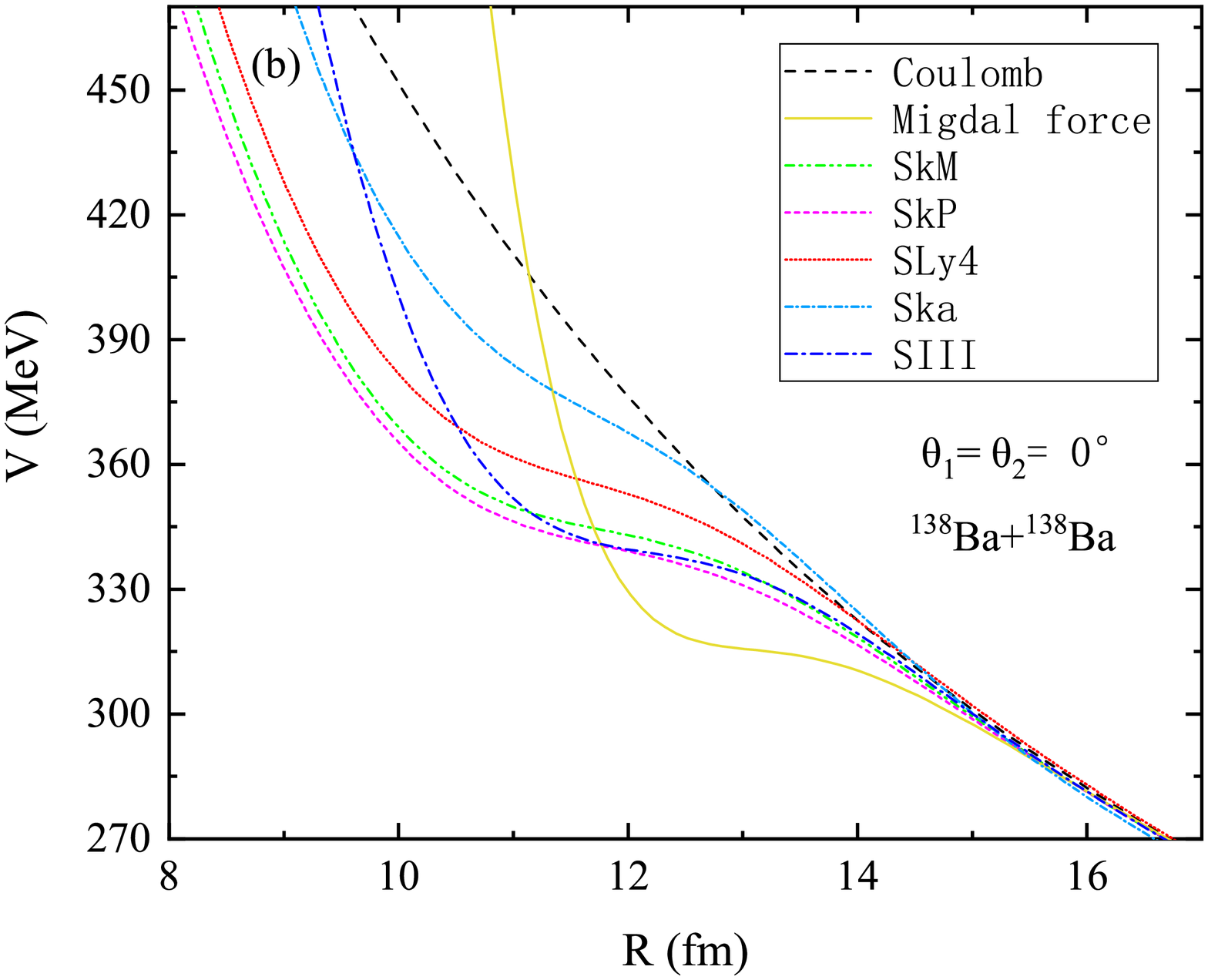}
\caption{Comparison of the nucleus-nucleus potentials from different Skyrme forces and Migdal force in collisions of $^{70}$Zn+$^{208}$Pb (left panel) and $^{138}$Ba+$^{138}$Ba (right panel), respectively.}
\end{figure*}
%%%%%%%%%%%%%%%%%%%%%%%%%%%%%%%%%%%%%%%%%%%%%%%%%%%%%%%%%%%%%%%

%%%%%%%%%%%%%%%%%%%%%%%%%%%%%%%%%%%% figure 3 %%%%%%%%%%%%%%%%%%%%%%
\begin{figure*}
	\includegraphics[width=16 cm]{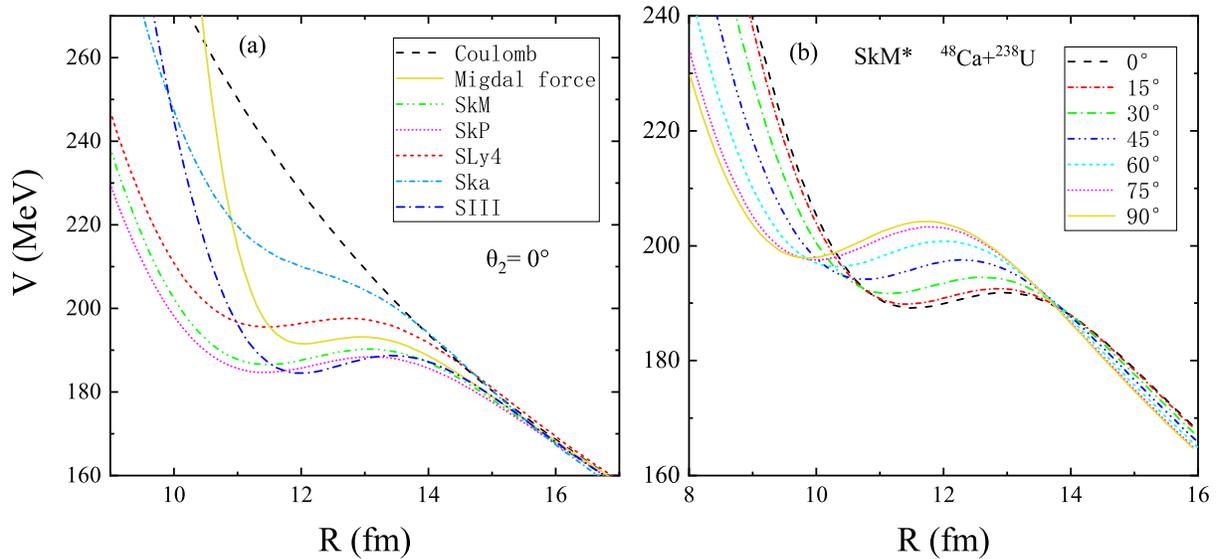}
	\caption{The nucleus-nucleus potentials with different Skyrme forces (left panel)  and the collision orientation with SkM$^{\ast}$ (right panel) in the reaction of $^{48}$Ca+$^{238}$U.}
\end{figure*}
%%%%%%%%%%%%%%%%%%%%%%%%%%%%%%%%%%%%%%%%%%%%%%%%%%%%%%%%%%%%%%%

\subsection{2.2  Formation of superheavy nucleus in fusion-evaporation reaction}

The formation of superheavy nuclei in massive fusion reaction is complicated and associated with the nucleon transfer, shape evolution, neck formation, relative motion energy and angular momentum dissipation. In the dinuclear system model, the density profiles of colliding nuclei are taken to be frozen and the neck dynamics is not taken into account. The nucleon transfer is coupled to the relative degrees of freedom via a set of master equations by the potential energy surface (PES). The PES is given by
\begin{equation}
U(R,\{\alpha\}) = Q_{gg}(Z_{1},N_{1}) + V(R,\{\alpha\})
\end{equation}
with $Q_{gg}(Z_{1},N_{1}) = B(Z_{1},N_{1})+B(Z_{2},N_{2}) - B(Z_{com},N_{com})$.The $B(Z_{i},N_{i}) (i=1,2)$ and $B(Z_{com},N_{com})$ are the negative binding energies of the fragment $(Z_{i},N_{i})$ and the compound nucleus $(Z_{com},N_{com})$, respectively. The binding energies were obtained from the calculation of the finite-range liquid-drop model \cite{Pm95}. It should be noted that the more precision mass table was proposed by Wang \emph{et al.} with the macroscopic-microscopic approach \cite{Wa15}. The symbol $\{\alpha\}$ denotes the quantities of $Z_{1}, N_{1}, \theta_{1}, \theta_{2}, \beta_{1}, \beta_{2}$. The $\beta_{i}$ represent the quadrupole deformations of two DNS fragments at ground state. The $\theta_{i}$ denote the angles between the collision orientations and the symmetry axes of deformed nuclei. The nucleus-nucleus potential between fragments $(Z_{1},N_{1})$ and $(Z_{2},N_{2})$ includes the nuclear and Coulomb interaction. In the calculation, the distance $R$ between the centers of the two fragments is chosen to be the value at the touching configuration, in which the DNS is assumed to be formed. The tip-tip orientation is chosen in the calculation of the SHN cross section, which manifests the elongation shape along the collision direction and is favorable for the nucleon transfer. The nuclear structure effects, i.e., shell effect, odd-even etc, are included in the binding energy. It should be noted that the dynamical deformation is not implemented into the binding energy.

The DNS model has been applied to the the quasi-fission and fusion dynamics, multinucleon transfer reactions and deep inelastic collisions, in which the dissipation of relative motion and rotation of colliding system into the internal degrees of freedom is assumed at the touching configuration. The DNS system evolves along two main degrees of freedom to form a compound nucleus, namely, the radial motion via the decay of DNS and the nucleon transfer via the mass asymmetry $\eta=(A_{1}-A_{2})/(A_{1}+A_{2})$ or the charge asymmetry $\eta_{Z}=(Z_{1}-Z_{2})/(Z_{1}+Z_{2})$ \cite{Ad20b,Fe09b}. In accordance with the temporal sequence, the system undergoes the capture by overcoming the Coulomb barrier, the competition of quasi-fission and complete fusion by cascade nucleon transfer, and the formation of cold residue nuclide by evaporating $\gamma$-rays, neutrons, light charged particles and binary fission. The production cross section of the superheavy residue is estimated by the sum of partial wave with the angular momentum $J$ at incident center of mass energy $E_{c.m.}$ as,
\begin{eqnarray}
\sigma_{ER}(E_{c.m.}) = && \frac{\pi\hbar^{2}}{2\mu E_{c.m.}}\sum^{J_{max}}_{J=0}(2J+1)T(E_{c.m.},J)     \nonumber\\
&& \times P_{CN}(E_{c.m.},J)W_{sur}(E_{c.m.},J)
\end{eqnarray}
Here, $T(E_{c.m.},J)$ is the penetration probability and given by the Hill-Wheeler formula and a Gaussian-type barrier distribution \cite{15zq}. The distribution function is taken as the Gaussian form $f(B) = \frac{1}{N} exp[-((B-B_{m})/\Delta)^{2}]$, with the normalization constant satisfying the unity relation $\int f(B)dB=1$. The quantities $B_{m}$ and $\Delta$ are evaluated by $B_{m}=(B_{C} + B_{S})/2$ and $\Delta = (B_{C} - B_{S})/2$, respectively. The $B_{C}$ and $B_{S}$ are the Coulomb barrier at waist-to-waist orientation and the minimum barrier by varying the quadrupole deformation of the colliding partners. The fusion probability $P_{CN}$ is described by the DNS model and taking into account the competition of the quasi-fission and fission of the heavy fragment \cite{Fe09b}, in which the nucleon transfer is described by solving a set of microscopically derived master equations by distinguishing protons and neutrons. The survival probability $W_{sur}$ is calculated with the Weisskopf statistical theory \cite{Ch17}, in which the decay of compound nucleus formed in the fusion reaction is cooled by evaporating $\gamma$ rays and light particles including neutrons, protons, $\alpha$ in competition with binary fission.

\section{III. Results and discussion}

The nucleus-nucleus potential is of significance in the low-energy heavy-ion collisions, i.e., the quasifission dynamics, fusion-fission reaction, fusion-evaporation for synthesizing the heavy or superheavy nuclei. The energy density function approach manifests a bridge between the nucleon-nucleon force and nuclear equation of state. The effective Skyrme force is expected for a unified description of the massive fusion reaction and density dependence of nuclear matter. The potential energy surface governs the nuclear dynamics in the fusion-evaporation and fusion-fission reactions. Shown in Fig. 4 is the PES as functions of the mass asymmetry and the center of mass distance of DNS fragments in the reaction of $^{48}$Ca+$^{238}$U calculated with the Skyrme forces SkP and SIII, respectively. The hard equation of state by SIII with the incompressible modulus of 352 MeV leads to the rapidly increasing when the two fragments approaching because of the more repulsive nucleon-nucleon force at the overlapping density above the saturation density. It has been known that the nucleus-nucleus potential manifests the attractive interaction when the local density below the normal nuclear density and the appearance of the Coulomb barrier or platform in the potential configuration. The mass symmetric fragments with $\eta\rightarrow 0$ have the large positive interaction potential and negative $Q_{gg}$. The competition results in the bump structure in the PES along the mass asymmetry degree of freedom. On the other hand, the collision orientation affects the interaction potential and also the PES for the deformed binary fragments. The nose-nose orientation is taken in the calculation. It is noticed that the density profile is assumed to be fixed in the calculation of PES, which is different with the multidimensional adiabatic potential used in Langevin equations \cite{Za07}.

%%%%%%%%%%%%%%%%%%%%%%%%%%%%%%%%%%%% figure 4 %%%%%%%%%%%%%%%%%%%%%%
\begin{figure*}
	\includegraphics[width=8 cm]{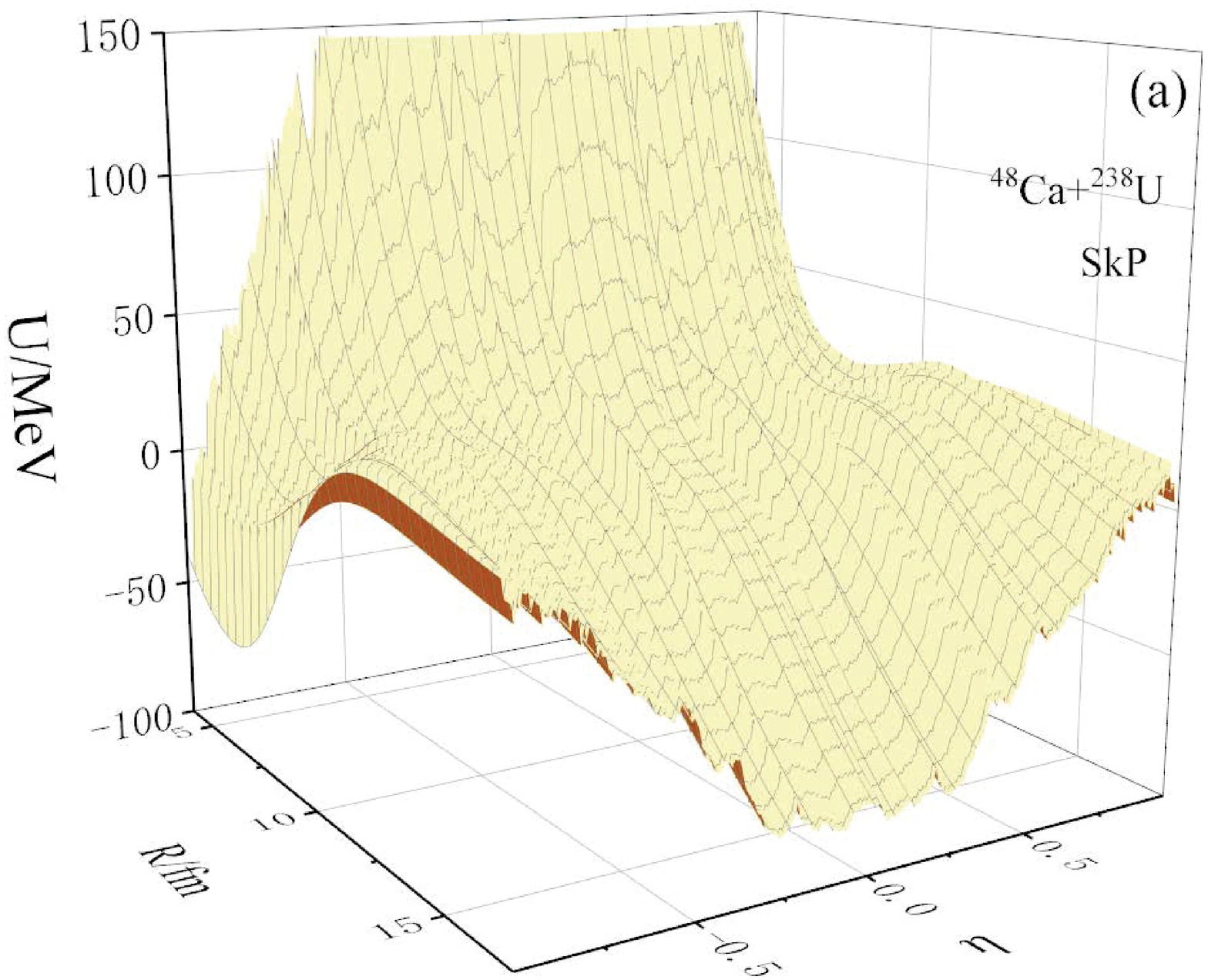} \quad \includegraphics[width=8 cm]{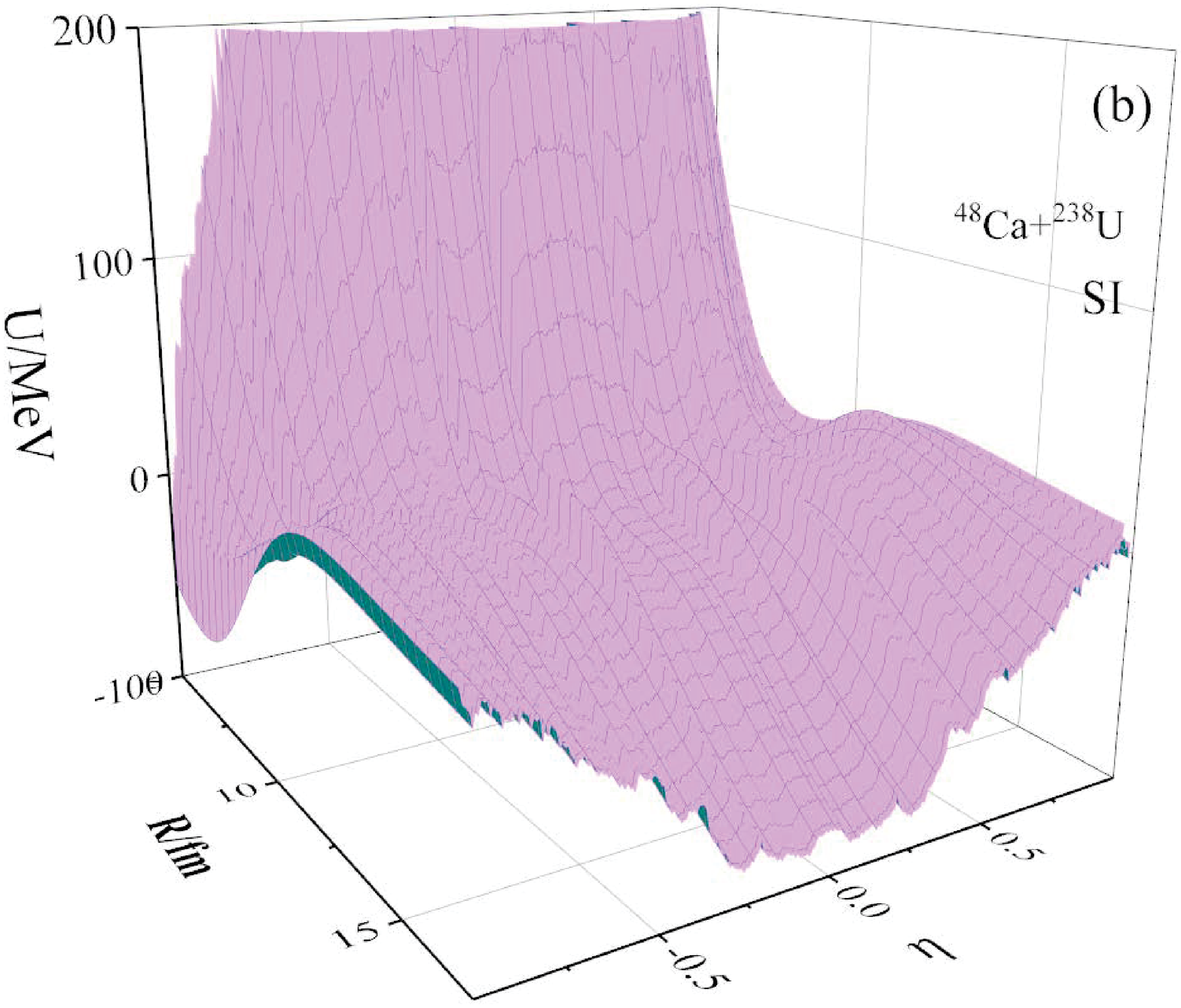}
	\caption{Potential energy surface as functions of the mass asymmetry and the center of mass distance in the reaction of $^{48}$Ca+$^{238}$U calculated with the Skyrme forces SkP and SIII, respectively. }
\end{figure*}
%%%%%%%%%%%%%%%%%%%%%%%%%%%%%%%%%%%%%%%%%%%%%%%%%%%%%%%%%%%%%%%

%%%%%%%%%%%%%%%%%%%%%%%%%%%%%%%%%%%% figure 5 %%%%%%%%%%%%%%%%%%%%%%
\begin{figure}
\includegraphics[width=8 cm]{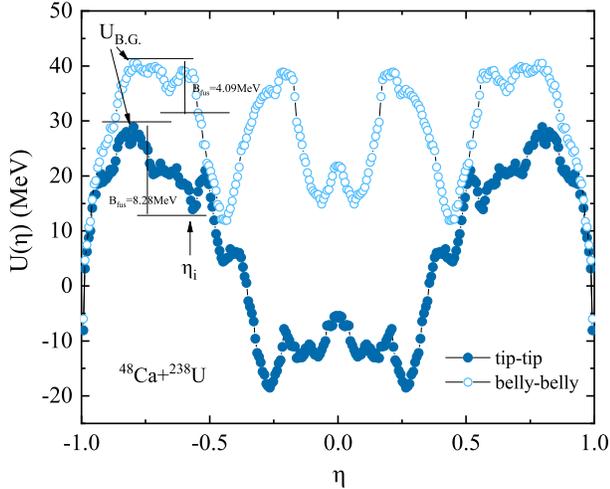}
\caption{The driving potentials in the tip-tip and belly-belly collisions for the reaction $^{48}$Ca + $^{238}$U with the force SkM$^{\ast}$.}
\end{figure}
%%%%%%%%%%%%%%%%%%%%%%%%%%%%%%%%%%%%%%%%%%%%%%%%%%%%%%%%%%%%%%%

The driving potential is taken the PES of DNS fragments at the touching configuration, namely, the minimal value in the potential pocket. The structure of the driving potential influences the quasifission yields, fusion-fission products, isotopic distribution in the multinucleon transfer reaction, SHN formation etc. The dependence of collision orientation of the driving potential in the reaction of $^{48}$Ca + $^{238}$U with the force SkM$^{\ast}$ is shown in Fig. 5. The entrance system is indicated at the position $\eta_{i}$, which is located the minimum in the tip-tip collision and maximal value of the driving potential in the belly-belly orientation at the beginning of nucleon transfer. The driving potential exhibits the symmetric structure with the mass asymmetry and is chosen with the lower potential between transferring a proton and neutron at the fixed collision angle and quadrupole deformations of DNS fragments. The distribution probability is obtained with the driving potential by solving a set of master equations. The fusion probability is counted via the left side of B.G. (Businaro-Gallone) point. The inner fusion barrier is estimated by the difference of B.G. position and entrance point. The diffusion to the right side from the entrance position with $\eta_{i}=-0.664$ leads to the formation of quasifission products. The bump in the belly-belly collision prevents the quasifission reaction and is favorable for the compound nucleus formation. The local minimum in the spectrum is caused from the shell correction on the binding energy.

Accurate estimation of the driving potential is of significance in the calculation of fusion probability, which is complicated and not well understood up to now. The fusion hindrance after overcoming the Coulomb barrier in colliding partners leads to the lowering of fusion probability to form a compound nucleus. The interaction time, shape evolution, dissipation of relative motion energy and angular momentum, coupling of nucleon transfer to the dynamical deformation, friction coefficient, mass tensor, neck dynamics etc, influence the compound nucleus formation in the heavy-mass fusion reactions. Shown in Fig. 6 is a comparison of SHN production in the reaction of $^{48}$Ca+$^{238}$U calculated by the Skyrme energy density functional with SkM$^{\ast}$ and the double folding approach with the Migdal force. It is pronounced that the available data from Dubna \cite{Og04} are nicely reproduced with the density functional approach. The Migdal force is usually taken in the calculation of driving potential in the DNS model \cite{15zq}. It should be noticed that the frozen density is taken into account in the nuclear potential for both methods.
It is obvious that the broader and more depth potential pocket is formed with the Skyrme force SkM$^{\ast}$ as shown in Fig. 3, which reduces the inner fusion barrier and enhances the production cross section of SHN. Similarly, the Skyrme parameters with the stiff equation of state lead to the decrease of evaporation residue cross section owing to the narrower potential pocket. The Pauli principle and nuclear equation of state in the formation of SHN is still interesting and needs to be investigated.

%%%%%%%%%%%%%%%%%%%%%%%%%%%%%%%%%%%% figure 6 %%%%%%%%%%%%%%%%%%%%%%
\begin{figure}
	\includegraphics[width=8 cm]{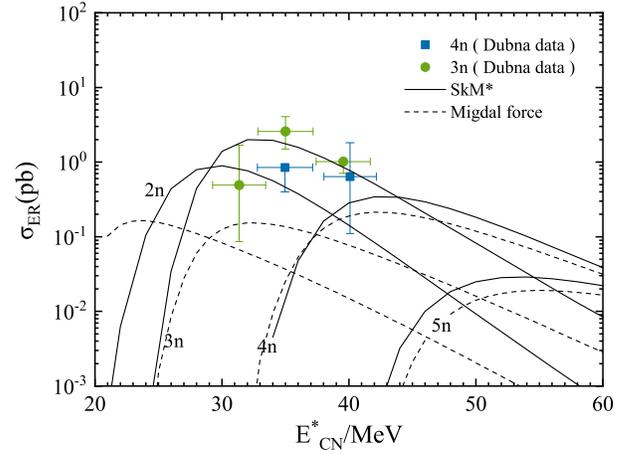}
	\caption{Comparison of evaporation residue excitation functions in the reaction of $^{48}$Ca+$^{238}$U for producing copernicium.}
\end{figure}
%%%%%%%%%%%%%%%%%%%%%%%%%%%%%%%%%%%%%%%%%%%%%%%%%%%%%%%%%%%%%%%

\section{IV. Conclusions}

In summary, the nucleus-nucleus potential is calculated with the Skyrme energy density functional, which is associated with the nuclear equation of state via the parameters SkM, SkM$^{\ast}$, SkP, SIII, Ska and SLy4. A broad and deep potential pocket is obtained with decreasing the incompressible modulus of nuclear matter with the Skyrme forces, which is favorable for the DNS formation. The repulsive nuclear potential is obvious with SIII when the overlap density above the saturation density. There are roughly 20 MeV difference for the Coulomb barriers of various Skyrme forces. The density profiles of projectile and target nuclides depend on the collision orientation for the deformed nuclei and lead to the difference of the interaction potential. The potential rapidly increases with reducing the center of mass distance of colliding partners after the touching configuration. The potential energy surface is associated with the nucleon-nucleon forces and manifests the structure effect. The Dubna data for synthesizing the copernicium (Z=112) in the reaction of $^{48}$Ca+$^{238}$U are nicely reproduced by the energy density functional with SkM$^{\ast}$, which corresponds to the soft nuclear equation of state (K=217 MeV) and the wider potential pocket in comparison with the usual double-folding approach with Migdal force.

\textbf{Acknowledgements}

This work was supported by the National Natural Science Foundation of China (Projects No. 12175072 and No. 11722546) and the Talent Program of South China University of Technology (Projects No. 20210115).

\section{Appendix}
The Hamiltonian of the $N$-body system is written as
\begin{eqnarray}
 \hat{H}=\sum_{i}\hat{t}_{i}+\sum_{i<j}\hat{t}_{ij}^{(2)}+\sum_{i<j<k}\hat{t}_{ijk}^{(3)}.
\end{eqnarray}
So the expectation value of the Hamiltonian in a Slater determinant $|HF>$ is given by
\begin{eqnarray}
 E&&=<HF|\hat{H}|HF>           \nonumber \\
 &&=\sum_{i}<i|\hat{t}|i>+\frac{1}{2}\sum_{ij}<ij|\bar{v}^{(2)}|ij>     \nonumber \\
 &&+\frac{1}{6}\sum_{ijk}<ijk|\bar{v}^{(3)}|ijk>.
\end{eqnarray}
The total energy of $N-$nucleon system is also expressed by the energy-density functional as
\begin{eqnarray}
 E=\int \varepsilon(\textbf{r}) d\textbf{r}.
\end{eqnarray}
The density function can be obtained by summing up all possible single-particle states $\phi_{i}(\bm{r},\sigma,q)$ with the spin $\sigma=\pm\frac{1}{2}$ and isospin symbols $q=\pm\frac{1}{2}$ ($q=\frac{1}{2}$ for neutron and $q=-\frac{1}{2}$ for proton). The nucleon density, kinetic energy density and spin-orbit current density can be obtained with the state function as
\begin{eqnarray}
 \rho_{q}(\bm{r})=\sum_{i,\sigma}|\phi_{i}(\bm{r},\sigma,q)|^{2},      \\
 \tau_{q}(\bm{r})=\sum_{i,\sigma}|\bigtriangledown\phi_{i}(\bm{r},\sigma,q)|^{2}
 \end{eqnarray}
 and
\begin{eqnarray}
 \bm{J}_{q}(\bm{r})=-i\sum_{i,\sigma,\sigma^{'}}\phi_{i}^{*}(\bm{r},\sigma,q)[\bigtriangledown\phi_{i}^{*}(\bm{r},\sigma^{'},q)\times<\sigma|\bm{\hat{\sigma}}|\sigma^{'}>], \nonumber \\
\end{eqnarray}
respectively.
For example, the first term with $t_{0}$ and $x_{0}$ is obtained by including the exchange operator to the matrix element as
\begin{eqnarray}
 \bar{v}_{k_{1},k_{2},k_{3},k_{4}} && = v_{k_{1},k_{2},k_{3},k_{4}}-v_{k_{1},k_{2},k_{4},k_{3}}  \nonumber \\
 &&=<k_{1}k_{2}|\hat{v}|k_{3}k_{4}>-<k_{1}k_{2}|\hat{v}|k_{4}k_{3}>  \nonumber \\
 &&=<k_{1}k_{2}|v(1-\hat{P_{M}}\hat{P_{\sigma}}\hat{P_{\tau}})|k_{3}k_{4}>
\end{eqnarray}
with the spatial exchange operator $P_{M}$, the spin exchange $P_{\sigma}$ and the isospin exchange $P_{\tau}$. The energy is calculated by substituting the state variables $(i,\sigma, q)$ for $k$ and leads to
\begin{eqnarray}
 E_{0}&& = \frac{1}{2}\sum_{ij,\sigma\sigma^{'},qq^{'}}<ij\sigma\sigma^{'}qq^{'}|t_{0}(1+x_{0}\hat{P}_{\sigma})\delta(\bm{r}_{1}-\bm{r}_{2})\nonumber \\
 && \times (1-\hat{P_{M}}\hat{P_{\sigma}}\hat{P_{\tau}})|ij\sigma\sigma^{'}qq^{'}>,
\end{eqnarray}
With the properties of the $\delta$ function and the exchange operator $P_{M}$, it has the similar isospin index $\hat{P_{\tau}}\rightarrow\delta_{qq^{'}}$. For the spin operator $\hat{P_{\sigma}} = \frac{1}{2}(1+\hat{\sigma_{1}}\cdot\hat{\sigma_{2}})$, one can obtain
\begin{eqnarray}
 &&E_{0} = \frac{1}{2}t_{0}\sum_{ij,\sigma\sigma^{'},qq^{'}}<ij\sigma\sigma^{'}qq^{'}|\delta(\bm{r}_{1}-\bm{r}_{2})  \nonumber \\
&& \times[1+\frac{1}{2}x_{0}(1+\hat{\sigma_{1}}\cdot\hat{\sigma_{2}})]            \nonumber \\
&&\times[1-\frac{1}{2}\delta_{qq^{'}}(1+\hat{\sigma_{1}}\cdot\hat{\sigma_{2}})]|ij\sigma\sigma^{'}qq^{'}>   \nonumber \\
&&=\frac{1}{2}t_{0}\sum_{ij,\sigma\sigma^{'},qq^{'}}<ij\sigma\sigma^{'}qq^{'}|\delta(\bm{r}_{1}-\bm{r}_{2})  \nonumber \\
&& \times[1+\frac{1}{2}x_{0} - \frac{1}{2}\delta_{qq^{'}} - \frac{x_{0}}{4}\delta_{qq^{'}}(1+\hat{\sigma_{1}}\cdot\hat{\sigma_{2}})^{2}] |ij\sigma\sigma^{'}qq^{'}>.   \nonumber \\
\end{eqnarray}
By substituting the relation $(\hat{\sigma_{1}}\cdot\hat{\sigma_{2}})^{2} = 3 - 2\hat{\sigma_{1}}\cdot\hat{\sigma_{2}}$,
the state wave function is implemented into the summation
\begin{eqnarray}
&& E_{0}  = \frac{1}{2}t_{0}\sum_{ij,\sigma\sigma^{'},qq^{'}}\int d^{3}r |\phi_{i}(\bm{r},\sigma,q)|^{2}
 |\phi_{j}(\bm{r},\sigma,q)|^{2}                                             \nonumber  \\
 && \times [1+\frac{1}{2}x_{0}-\delta_{qq^{'}}[x_{0}+\frac{1}{2}]]         \nonumber \\
 &&=\frac{1}{2}t_{0}\sum_{qq^{'}}\int d^{3}r\rho_{q}(\bm{r})\rho_{q^{'}}(\bm{r})
 [1+\frac{1}{2}x_{0}-\delta_{qq^{'}}[x_{0}+\frac{1}{2}]],      \nonumber \\
\end{eqnarray}
The energy-density functional is given by
\begin{eqnarray}
 \varepsilon_{0}(\bm{r})=\frac{t_{0}}{2} \left[ (1+\frac{1}{2}x_{0})\rho^{2}(\bm{r})-(x_{0}+\frac{1}{2})(\rho_{p}^{2}(\bm{r})+\rho_{n}^{2}(\bm{r}))\right]    \nonumber \\
\end{eqnarray}
with the total density $\rho=\rho_{n}+\rho_{p}$. Similarly, one can get the other terms in Eq. (13).

\end{document}